\documentclass[12pt]{revtex4}
\usepackage{amssymb}
\usepackage{latexsym}
\usepackage{epsfig}
\begin{document}                             

\title{ Constructing warm  inflationary model in brane-antibrane system
 }

\author{M. R. Setare $^{1}$\footnote{rezakord@ipm.ir}, A.Sepehri $^{2}$\footnote{alireza.sepehri@uk.ac.ir}, V.Kamali $^{3}$ \footnote{Vkamali@basu.ac.ir
}}
\address{$^1$ Department of Science, Campus of Bijar,
University of Kurdistan, Bijar, Iran.\\ $^2$  Faculty of Physics, Shahid Bahonar University, P.O. Box 76175, Kerman, Iran.\\ $^3$ Department of Physics, Faculty of Science, Bu-Ali Sina University, Hamedan, 65178, Iran.
}

\begin{abstract}
Recently, various observational data predict a possibility that inflation may naturally occur in a warm region. In this scenario, radiation is produced during the inflation epoch and reheating is avoided. The main question arises that what is the origin of warm inflation in 4D universe? We answer to this question in brane-antibrane system. We propose a model that allows all cosmological parameters like the scale factor a, the Hubble parameter H and phantom energy density depend on the equation of state parameter in transverse dimension between two branes. Thus, an enhancement in these parameters can be a signature of some evolutions in extra dimension. In our model, the expansion of 4D universe is controlled by the separation distance between branes and evolves from non-phantom phase to phantom one. Consequently, phantom-dominated era
of the universe accelerates and ends up in big-rip singularity. Also, we show that as the tachyon potential increases, the effect of interaction between branes on the 4D universe expansion becomes systematically more effective, because
at higher energies there exists more channels for flowing energy from extra dimension to other four dimensions. Finally, we test our model against WMAP and Planck data and obtain the ripping time. According to experimental data, $N\simeq 50$ case leads to $n_{s}\simeq 0.96$, where \emph{N} and $n_{s}$ are the number e-folds and the spectral index respectively. This standard case may be found in $0.01 < R_{Tensor-scalar } < 0.22$, where $R_{Tensor-scalar }$ is the tensor-scalar ratio. At this point, the finite time that Big Rip singularity occurs is $t_{rip}=33(Gyr)$.

\end{abstract}

 \maketitle
\section{Introduction}
One main problem of the inflation theory is how to attach the universe to the end of the
inflation epoch. An interesting solution of this problem is the study of inflation in the
context of warm inflation scenario \cite{m1}. In this model, radiation is produced during the inflation epoch and reheating is avoided.
Until now, the warm inflationary universe has been discussed in many papers \cite{m2,m3,m4,m5,m6}. Specially, the effect of interaction between branes on the warm inflation has been considered in some scenarios \cite{m7,m8,m9}. For example, some authors showed that for D3-$\bar{D3}$ branes with additional flavor D7-branes, inflation may naturally occur in a warm regime. They explained that  by using a Coulomb-like or quadratic hybrid potential, a
sufficient number of e-folds may be calculated for perturbative couplings and $O(10-10^{4})$ D3-branes. This was successful realizations of warm
inflation and lead to a sufficiently long
period of accelerated expansion even for the steep potentials
dominating brane-antibrane interactions in flat
space. Furthermore, their model  predicted
interesting observational signatures such as negligible tensor-to-scalar ratio \cite{m7}.
Some other authors discussed  that the unstable tachyonic field which arises from the motion of D3 brane in the background of k coincident D5 branes can be the source of inflation in warm inflationary scenario. The whole process was considered in presence of a radiation bath. In this model, the dissipative effects had important role and hence the whole dynamics was analyzed in terms of the
dissipative parameter. Also, the slow-roll parameters and the cosmological observables were evaluated by considering the
fact that the dissipative parameter is a function of the tachyonic field. This field was given in terms of the radius of the ring produced by one D3 and k number of D5 branes \cite{m8}. In another research, the study of warm inflation model as a mechanism that gives an end for vector inflation theory is motivated some authors to consider the warm vector inflation model \cite{m10}. In their scenario, the general conditions for inflation era and end of this epoch were obtained.

 Recently, some investigators \cite{m9} examined the warm inflationary model with observational data and showed that the prediction of this model is consistent with Planck results \cite{m11}. They explained that when the temperature of the radiation
bath is bigger or equal to Hubble parameter ($T\geq H$) and  non-trivial inflaton occupation numbers are sustained during inflation, the
amplitude of scalar curvature fluctuations can be significantly enhanced, whereas tensor perturbations are generically
unaffected due to their weak coupling to matter
fields. This generically reduces the tensor-to-scalar ratio
with respect to conventional models, changes
the tilt of the scalar power spectrum and yields a modified consistency relation for warm inflation. They discussed that as an example, the quartic chaotic potential is in very good agreement with the
Planck results for nearly-thermal inflaton
fluctuations, whereas it seems to be ruled out in a
cold scenario. They also consider other simple models that are in agreement with the Planck data \cite{m11} within a renormalizable model of warm inflation.

 Now, the question arises that what is the origin of warm inflationary model in 4D universe? We answer to this question in brane-antibrane system. In this system, it might be reasonable to ignore the tachyon in the
ultraviolet where the flavor branes and antibranes are well separated and the tachyon is
small, it is likely to condense and acquire large values in the infrared where the branes
meet. In this condition, inflation of universe is controlled by tachyon  potential between branes  and evolves from non phantom phase to phantom one. The tachyonic phantom
energy density grows without bound and  becomes infinite in finite time, dominates all other forms of
energy, such that the gravitational repulsion  and finally  brings our brief epoch of universe to
an end. When the universe approaches this future singularity; the Milky Way, solar system, Earth, and ultimately the
molecules, atoms, nuclei, and nucleons of which we are composed will be ripped by the infinite phantom energy
in a "Big Rip" \cite{m13}. This singularity is  characterized by divergences in the scale factor a, the Hubble parameter H  and its
time-derivative $\dot{H}$  at the finite future $t = t_{rip}$.

The outline of the paper is as the following.  In section \ref{o1}, we construct warm inflationary model of universe in brane-antibrane system and show that all cosmological parameters depend on the separation distance between two branes. Also, in this section, we consider Big Rip singularity and calculate the relation between ripping time and separation  between branes in extra dimension. In section \ref{o2}, we test our model against the observational data from Planck and WMAP collaborations \cite{m11,m12,m27} and obtain the ripping time . The last section is devoted to summary and conclusion.

\section{ The warm  inflationary model of universe in D3/$\bar{D3}$ system}\label{o1}
Previously, the dynamical behavior of warm brane-antibrane inflation in the era that the Big Rip singularity
will not occur was considered \cite{m7,m8}. In this section we will enter the effects of this singularity on the results of the derivation of scale factors, Hubble parameter and other important parameters in warm  model model of cosmology. We will show that these parameters depend on the ripping time and are influenced by separation distance of branes in extra dimension. In our model, equation of state parameter in 4D universe may change due to flowing energy from extra dimension and decrease from higher values of -1 (non-phantom phase) to lower values (phantom one).

  The D3/$\bar{D3}$ model is formulated by placing $\bar{D3}$-branes into the D3-brane background with the following metric \cite{m14,m15,m16,m17,m18}:
 \begin{eqnarray}
&& ds^{2}=( 1 + \frac{R^{4}}{y^{4}} )^{-\frac{1}{2}}\eta_{\mu\nu}dx^{\mu}dx^{\nu} + ( 1 + \frac{R^{4}}{y^{4}} )^{\frac{1}{2}}( dy^{2} + y^{2}d\Omega_{5}^{2} ).
\label{Q1}
\end{eqnarray}
Here
\begin{eqnarray}
&& R_{4}=4\pi g_{s}N\acute{\alpha}^{2},
\label{Q2}
\end{eqnarray}
where $\lambda = g_{s}N = g_{YM}^{2}$ is the 't Hooft coupling, N the number of D3 branes and $\acute{\alpha}$ the inverse string tension ($\acute{\alpha}$ = $l_{s}^{2}$, $l_{s}$ string length). Also, $\eta_{\mu\nu}$ is the standard 3+1 dimensional Minkowski
metric, the $x_{\mu}$ are the coordinates on the stack of D3 branes and the $y$ denotes the spatial
coordinate perpendicular to the brane.

 To construct four dimensional universe, we consider a set of  D3-$\overline{D3}$-brane pairs in the above background which are placed
at points $y_{1} = l/2$ and $y_{2} = -l/2$ respectively so that the
separation between the brane and antibrane is l. For the simple case of a single D3-$\overline{D3}$-brane pair with open string tachyon, the action is\cite{m19}:
 \begin{eqnarray}
&& S=-\tau_{3}\int d^{9}\sigma \sum_{i=1}^{2} V(T,l)e^{-\phi}(\sqrt{-det A_{i}})\nonumber \\&& (A_{i})_{ab}=(g_{MN}-\frac{T^{2}l^{2}}{Q}g_{My}g_{yN})\partial_{a}x^{M}_{i}\partial_{b}x^{M}_{i}
+F^{i}_{ab}+\frac{1}{2Q}((D_{a}T)(D_{b}T)^{\ast}+(D_{a}T)^{\ast}(D_{b}T))\nonumber \\&&
+il(g_{ay}+\partial_{a}y_{i}g_{yy})(T(D_{b}T)^{\ast}-T^{\ast}(D_{b}T))+
il(T(D_{a}T)^{\ast}-T^{\ast}(D_{a}T))(g_{by}+\partial_{b}y_{i}g_{yy}).
\label{Q3}
\end{eqnarray}
where
  \begin{eqnarray}
&& Q=1+T^{2}l^{2}g_{yy},
\nonumber \\&& D_{a}T=\partial_{a}T-i(A_{2,a}-A_{1,a})T, V(T,l)=g_{s}V(T)\sqrt{Q},
\nonumber \\&& e^{\phi}=g_{s}( 1 + \frac{R^{4}}{y^{4}} )^{-\frac{1}{2}},
\label{Q4}
\end{eqnarray}
,$\phi$, $A_{2,a}$ and $F^{i}_{ab}$ are dilaton field, the gauge fields and field strengths on the world-volume of the non-BPS brane, T is the tachyon field, $\tau_{3}$ is the brane tension  and V (T) is
the tachyon potential. The indices a,b denote the tangent directions of D-branes, while the indices M,N run over the background ten-dimensional space-time directions. The Dp-brane and the anti-Dp-brane are labeled by i = 1 and
2 respectively. Then the separation between these D-branes is defined by $y_{2} - y_{1} = l$. Also, in writing the above we are using the convention $2\pi\acute{\alpha}=1$.

 Let us consider the only t
dependence of the tachyon field T for simplicity and set the gauge fields to zero. In this case, the static gauge the action (\ref{Q3}) in the region that  $r> R$ simplifies to
  \begin{eqnarray}
S\simeq-\frac{\tau_{3}}{g_{s}}\int d^{4}x V(T)(\sqrt{D_{1,T}}+\sqrt{D_{2,T}})
\label{Q5}
\end{eqnarray}
where $D_{1,T} = D_{2,T}\equiv D_{T}$, $V_{3}=\frac{4\pi^{2}}{3}$ is the volume of
a unit $S^{3}$ and
 \begin{eqnarray}
D_{T} = 1 + \frac{l'(t)^{2}}{4}+
T'(t)^{2} + T(t)^{2}l(t)^{2}
\label{Q6}
\end{eqnarray}
where the 'prime'  denotes a derivative w.r.t. its argument t.

The equations of motion obtained from this action are:
 \begin{eqnarray}
(\frac{1}{\sqrt{D_{T}}}T'(t)\acute{)}=\frac{1}{\sqrt{D_{T}}}
[T(t)l(t)^{2}+\frac{V'(T)}{V(T)}(D_{T}-T'(t)^{2})]
\label{Q7}
\end{eqnarray}
 \begin{eqnarray}
(\frac{1}{\sqrt{D_{T}}}l'(t)\acute{)}=\frac{1}{\sqrt{D_{T}}}
[T(t)^{2}l(t)-\frac{V'(T)}{V(T)}l'(t)T'(t))]
\label{Q8}
\end{eqnarray}
Note that the 'prime' on V (T) denotes a derivative w.r.t. its argument T and not
a derivative w.r.t. time. To solve these equations, we should determine explicit form of V (T). There are several proposals for V (T) which satisfy these requirements \cite{m20},
although no rigorous derivation exists. In view of this, in the following analysis we will avoid using any specific expression for V (T), except when needed for explicit numerical calculations. It will, however, be necessary for us to specify the asymptotic form of the potential for large T. A potential which satisfies above equations and has been used in most of papers is \cite{m21,m22,m23}:
 \begin{eqnarray}
V(T)=\frac{\tau_{3}}{cosh\sqrt{\pi}T}
\label{Q9}
\end{eqnarray}
  Employing this potential, we can solve equations (\ref{Q7} and \ref{Q8}). Approximate solutions of these equations for old universe are:
\begin{eqnarray}
&& T(t)\sim\frac{T_{+}}{(t-t_{rip})}(1+\frac{1}{l_{0}(t-t_{rip})})e^{-l_{0}(t-t_{rip})}+\frac{T_{-}}{(t-t_{rip})}(1-\frac{1}{l_{0}(t-t_{rip})})e^{l_{0}(t-t_{rip})}\nonumber \\&& l(t)\sim l_{1}(t-t_{rip})+.....
\label{Q10}
\end{eqnarray}
 where $l_{0}$ is the maximum separation between the brane and antibrane and $T_{+}$ and $l_{1}$ are positive constants which will be determined in future calculations. These equations show that when two branes are very distance from each other (t=0), tachyon field is very small, whereas moving the  branes towards each other, the value of this field increases and  becomes infinity at $t = t_{rip}$.

 The effective potential for the tachyon can be obtained from the action (\ref{Q5}) by setting $T'=h'=0$. It is:
\begin{eqnarray}
&& V_{eff}(T)\sim sech\sqrt{\pi}T\sqrt{1 + \frac{l'(t)^{2}}{4}+
T'(t)^{2} + T(t)^{2}l(t)^{2}}\rightarrow \nonumber \\&&
V_{eff}(T)\sim sech\sqrt{\pi}(\frac{T_{+}}{(t-t_{rip})})\sqrt{1 + \frac{T_{+}}{(t-t_{rip})^{4}}}
\label{Q11}
\end{eqnarray}
This potential indicates that when the  branes-antibranes become close to each other, the interaction potential increases and at higher energies there exists more channels for string tachyon production in this system; consequently, the effect of string tachyon on the universe inflation becomes systematically more effective.

Let us now discuss the warm  inflationary model of universe in D3/$\bar{D3}$ system. For this, we
need to compute the contribution of the flavour brane-antibrane system to the four-
dimensional universe energy momentum tensor. The energy momentum tensor is obtained from action by calculating its functional
derivative w.r.t. the background ten-dimensional metric $g_{MN}$. The precise relation
is $T^{MN} = \frac{2}{\sqrt{-det g}}\frac{\delta S}{\delta g_{MN}}$. We get, \cite{m19},
 \begin{eqnarray}
&& T^{ab}_{i}=-V(T)\sqrt{D_{T}}g^{ab}, a,b=0,1,2,3 \nonumber \\&&
T^{yy}_{i}=-V(T)\frac{1}{\sqrt{D_{T}}}
(T^{2}l^{2}+\frac{\acute{l}^{2}}{4})
\label{Q12}
\end{eqnarray}
This equation shows that with decreasing the separation between D3 and $\overline{D3}$ branes, tachyon energy-momentum tensors increase. This is because that when  branes and antibranes are well separated, tachyon is small, however when two branes are close to each other, it achieves to large values.

The conservation law of energy-momentum relates the tensors calculated in brane-antibrane system with ones associated with the four dimensional universe with the following equation:
 \begin{eqnarray}
&& T^{\mu\nu} = \frac{2}{\sqrt{-det g}}\frac{\delta S}{\delta g_{MN}}\frac{\delta g_{MN}}{\delta g_{\mu\nu}} = T_{MN}\frac{\delta g_{MN}}{\delta g_{\mu\nu}},
\label{Q13}
\end{eqnarray}
where $T^{\mu\nu}$ is the energy-momentum tensor of 4D universe in ten dimensional space-time with the metric of the form:
\begin{equation}
ds^{2} = ds^{2}_{Uni,1} + ds^{2}_{Uni,2} + b^{2}dw^{2} + \omega^{2}d\Omega_{2},
\label{Q14}
\end{equation}
Here
 \begin{eqnarray}
&& ds^{2}_{Uni1} = ds^{2}_{Uni2} = -dt^{2} + a(t)^{2}(dx^{2} + dy^{2} + dz^{2}),
\label{Q15}
\end{eqnarray}
where $w$ is the extra space-like coordinates perpendicular to two universes  and scale factors $a(t)$ and
$b(t)$ are assumed to be functions of time only. In this model, we introduce two four dimensional  universe that interact with each other and form a binary system. To obtain the energy- momentum tensor in this system, we use of the Einstein's field equation in
presence of fluid flow  that reads as:
\begin{equation}
{R_{ij}} - \frac{1}{2}{{\mathop{\rm g}\nolimits} _{ij}}R = k{T_{ij}}.
\label{Q16}
\end{equation}
Setting the solution of this equation with the line element of Eq. (\ref{Q14}) in the conservation law  of energy-momentum tensor in Eq. (\ref{Q13}) and employing Eq. (\ref{Q12}) yield:
\begin{eqnarray}
&& kT_1^1 = \frac{{5\ddot a}}{a} + \frac{{{3{\dot a}^2}}}{{{a^2}}} + \frac{{5\dot a\dot
b}}{{ab}} + \frac{{\ddot b}}{b} = -V(T)\sqrt{D_{T}},  \nonumber\\
&& kT_2^2 =\frac{{5\ddot a}}{a} + \frac{{{3{\dot a}^2}}}{{{a^2}}} + \frac{{5\dot a\dot
b}}{{ab}} + \frac{{\ddot b}}{b} = -V(T)\sqrt{D_{T}}, \nonumber\\
&&  kT_3^3 = \frac{{5\ddot a}}{a} + \frac{{{3{\dot a}^2}}}{{{a^2}}} + \frac{{5\dot a\dot
b}}{{ab}} + \frac{{\ddot b}}{b} = -V(T)\sqrt{D_{T}}, \nonumber\\
&&  kT_4^4 =  \frac{{6\ddot a}}{a} + \frac{{6{{\dot a}^2}}}{{{a^2}}} = -V(T)\frac{1}{\sqrt{D_{T}}}(T^{2}l^{2}+\frac{\acute{l}^{2}}{4}), \nonumber\\
&& kT_{10}^{10} = \frac{{6{{\dot a}^2}}}{{{a^2}}} + \frac{{6\dot a\dot b}}{{ab}} = V(T)\sqrt{D_{T}} \label{Q17}
\end{eqnarray}
 The higher-dimensional stress-energy tensor will be assumed to be that of a perfect fluid and of the form:
\begin{equation}
T_i^j = {\mathop{\rm diag}\nolimits} \left[ { - p, - p, - p, - \bar{p}, - p, - p, - p, \rho } \right],
\label{Q18}
\end{equation}
where $\bar{p}$ is the pressure in the extra space-like dimension. In above equation, we are allowing the pressure in the extra
dimension to be different, in general, from the pressure in the 3D space. Hence, this stress-energy tensor describes a
homogeneous, anisotropic perfect fluid in ten dimensions. By adopting the metric ansatzs in  (\ref{Q1}) and (\ref{Q14}), the conservation law in (\ref{Q13}) and (\ref{Q17}) and the perfect fluid stress-energy tensor in (\ref{Q18}), the field equations are of the form:
\begin{equation}
- p = \frac{{5\ddot a}}{a} + \frac{{{3{\dot a}^2}}}{{{a^2}}} + \frac{{5\dot a\dot
b}}{{ab}} + \frac{{\ddot b}}{b} = -V(T)\sqrt{D_{T}},
\label{Q19}
\end{equation}
\begin{equation}
- \bar{p} = \frac{{6\ddot a}}{a} + \frac{{6{{\dot a}^2}}}{{{a^2}}} = -V(T)\frac{1}{\sqrt{D_{T}}}(T^{2}l^{2}+\frac{\acute{l}^{2}}{4}) ,
\label{Q20}
\end{equation}
\begin{equation}
 \rho = \frac{{6{{\dot a}^2}}}{{{a^2}}} + \frac{{6\dot a\dot b}}{{ab}} = V(T)\sqrt{D_{T}},
\label{Q21}
\end{equation}
where  we set the higher-dimensional
coupling constant equal to one, k = 1. The equations (\ref{Q19}),(\ref{Q20}) and (\ref{Q21}) help us to explain all properties of the current universe in terms of evolutions in brane-antibrane system. These equations constraint the pressure and density in our universe to tachyon fields and express that any increase or decrease in these parameters is due to tachyon potential in extra dimension.

 By taking derivative from the left side of Eq. (\ref{Q21}), we get the following equation:
\begin{equation}
\frac{2\dot{a}}{a}( \frac{5\ddot{a}}{a} - \frac{3\dot{a}^{2}}{a^{2}} - \frac{\dot{a}\dot{b}}{ab} + \frac{\ddot{b}}{b} ) + \frac{\dot{b}}{b}( \frac{2\ddot{a}}{a} - \frac{2\dot{a}\dot{b}}{ab} )
 = \frac{\dot{\rho}}{3}.
\label{Q22}
\end{equation}
Substituting  (\ref{Q19}) and (\ref{Q20}) in (\ref{Q22}), we obtain:
\begin{equation}
\frac{2\dot{a}}{a}( -\rho - p ) + \frac{\dot{b}}{b}( - \frac{\rho}{3} - \frac{\bar{p}}{3} )
 = \frac{\dot{\rho}}{3}.
\label{Q23}
\end{equation}
or
\begin{equation}
\dot{\rho} + 6\frac{\dot{a}}{a}( \rho + p) + \frac{\dot{b}}{b}( \rho + \bar{p}) = 0.
\label{Q24}
\end{equation}
 By defining $H = \frac{\dot{a}}{a}$ and $\bar{H} = \frac{\dot{b}}{b}$ as Hubble parameters in 4D universe and extra dimension, we can rewrite (\ref{Q21}) and (\ref{Q24}) as following:
\begin{equation}
6H^{2} + 6H\bar{H} = \rho.
\label{Q25}
\end{equation}
\begin{equation}
\dot{\rho} + 6H( \rho + p) + \bar{H}( \rho + \bar{p}) = 0.
\label{Q26}
\end{equation}
   Eq.(\ref{Q26}) is in fact conservation equation in ten dimensional space-time. This equation is analogous to standard 4D
FRW cosmology with $\bar{H}=0$ and $ds^{2}_{Uni2} =0$. We are now at a stage that can enter the properties of warm  inflation into  calculations. The dynamic of warm  inflation in spatially flat
ten dimensional space-time is presented by this equation:
\begin{eqnarray}
&& \dot{\rho} + 6H( \rho + p) + \bar{H}( \rho + \bar{p})  = - \Gamma \dot{T}^{2} . \label{Q33}
\end{eqnarray}
Also, with similar calculations we can obtain:
\begin{eqnarray}
\dot{\rho}_{\gamma} + (8H + \bar{H} )\rho_{\lambda}  = \Gamma \dot{T}^{2}. \label{Q333}
\end{eqnarray}
where $\rho_{\lambda}$
   is energy density of the radiation and $\Gamma$ is the dissipative coefficient. In the above
equations dot h.h means derivative with respect to cosmic time. Warm inflation is an example
of a phase transition with dissipative effect \cite{m1}. In supercool inflation model \cite{m24}, this effect
becomes important after the end of inflation (in reheating epoch,) but in warm inflation
model the interactions of inflaton with other fields are important during the inflationary
period.

 Relating the 3D and higher-dimensional pressures to the density  (p = $\omega \rho$, $\bar{p}$ = $\bar{\omega} \rho$) and employing Eq.(\ref{Q25}) in Eq.(\ref{Q33}) lead to following equation:
  \begin{eqnarray}
&& 6H\dot{H} + 3\dot{H}\bar{H} + 3H\dot{\bar{H}}  \nonumber\\
&& + 9(H^{3} + H^{2}\bar{H})( 1 + \omega ) \nonumber\\
&& + 9(H^{2}\bar{H} + H\bar{H}^{2})( 1 + \bar{\omega} ) = - \Gamma \dot{T}^{2}, \label{Q27}
\end{eqnarray}
where $\omega$ and $\bar{\omega}$ are equation of state parameters in 4D universes and extra dimension respectively. Notice that these parameters can
in general be time-dependent.  Also, the relation ($\bar{p}$ = $\bar{\omega} \rho$) and equations (\ref{Q20}) and (\ref{Q21}) yield:
\begin{equation}
\bar{H} = \frac{\dot{H} + H^{2}\bar{\omega}}{\bar{\omega} H}.
\label{Q28}
\end{equation}
Substituting Eq.(\ref{Q28}) in Eq.(\ref{Q27}) gives us the following relation:
\begin{eqnarray}
&& ( \bar{\omega}^{2} - 2\bar{\omega} - 3\bar{\omega}\omega +4 )H^{2}\dot{H}  \nonumber\\
&& + ( \bar{\omega}^{2} - 6\bar{\omega}\omega +4 )H^{4} \nonumber\\
&& + ( 1+ \bar{\omega} )\dot{H}^{2} - \bar{\omega}H\ddot{H} = - \Gamma \dot{T}^{2}. \label{Q29}
\end{eqnarray}
Solving  equations (\ref{Q19}, \ref{Q20}, \ref{Q21}, \ref{Q28}, \ref{Q29}) simultaneously  and performing some algebra for old universe, we find a consistent solution
for the  Hubble parameters and scale factors in 4D universes and extra dimension and also tachyon, separation between branes and dissipation coefficient in terms of ripping time:
\begin{eqnarray}
&& H = \frac{A}{t-t_{rip}}  \nonumber\\
&& a(t)=( t - t_{rip} )^{A}\nonumber\\
&& A = \frac{(\bar{\omega}^{2} - 2\bar{\omega} - 3\bar{\omega}\omega +4) \pm \sqrt{(\bar{\omega}^{2} - 2\bar{\omega} - 3\bar{\omega}\omega +4)^{2} - 4( \bar{\omega}^{2} - 6\bar{\omega}\omega +4 )( 1 - \bar{\omega} ) }}{2( \bar{\omega}^{2} - 6\bar{\omega}\omega +4 )}, \label{Q30}\\ \nonumber  \\
&& \bar{H} = \frac{B}{t-t_{rip}}  \nonumber\\
&& b(t) = ( t - t_{rip} )^{B}\nonumber\\
&& B = \frac{-1 + A\bar{\omega}}{\bar{\omega}}, \label{Q31}\\ \nonumber \\
&& T \sim \frac{A^{2} + AB}{(t-t_{rip})}(1+\frac{1}{l_{0}(t-t_{rip})})e^{-l_{0}(t-t_{rip})} +\frac{A^{2} - AB}{(t-t_{rip})}(1-\frac{1}{l_{0}(t-t_{rip})})e^{l_{0}(t-t_{rip})} \nonumber\\
&& l \sim (A^{2} + AB)^{\frac{9}{4}}(t-t_{rip}), \label{Q32}\\ \nonumber \\
&& \Gamma \sim \frac{A(t-t_{rip})^{4}}{(A^{2} + AB)}(1+\frac{1}{l_{0}(t-t_{rip})})^{-2}e^{2l_{0}(t-t_{rip})}. \label{Q34}
\end{eqnarray}
Equation (\ref{Q30}) shows that if A is smaller than zero, 3D scale factor and Hubble parameter at a certain time (t = $t_{rip}$) are infinity. So far, researchers have believed that if 3D equation of state parameter is lower than negative one, Dark energy is phantom energy and at Big Rip singularity, ingredients of which universe is made will be ripped; whereas our calculations indicate that occurrence of this singularity is related to both $\bar{\omega}$ and $\omega$.

As can be seen from (\ref{Q31}), the value of the scale factor is determined by two competing terms (A and $\bar{\omega}$). When $A<0$ and $\bar{\omega}>0$, both 4D universe and extra dimension encounter with a type of singularity. Also, if $\bar{\omega},A<0$ and $\bar{\omega}>\frac{1}{A}$ or $A>\frac{1}{\bar{\omega}}$, we are faced with similar situation; while if $\bar{\omega}<\frac{1}{A}$ or $A<\frac{1}{\bar{\omega}}$, Big Rip singularity won't happen in extra dimension. In this regard, another particular condition ($\bar{\omega},A>0$ and $\bar{\omega}<\frac{1}{A}$ or $A<\frac{1}{\bar{\omega}}$)may arise where this singularity occurs only in the extra dimension.

The equation (\ref{Q32}) has some interesting results which can be used to explain the reasons for occurance of Big Rip singularity in present era of universe. According to these calculations, when two branes are located at a large distance from
each other ($l=l_{0}$ and t=0), tachyon field is almost zero, while approaching the two together, the value
of this field increases and tends to infinity at $t = t_{rip}$. In this situation, brane-antibrane system disappears and consequently one singularity happens in our four dimensional universe. Another interesting point that comes out from this equation is that time of this singularity is  proportional to the initial distance between two branes.

The equation (\ref{Q34}) indicates that the dissipative coefficient decreases with time and shrink to zero at ripping time. Also, dissipative coefficient is controlled by both of equation of state parameters in four dimensional universes and extra dimension. This means that the warm inflation can be affected by phenomenological events in extra dimension and other universe.

Inserting the solutions in (\ref{Q30}, \ref{Q31}, \ref{Q34})
back into Eq.(\ref{Q333}), we obtain the energy density of radiation in terms of ripping time:
\begin{equation}
\rho_{\lambda}\sim \frac{1}{(t-t_{rip})^{8A+B}}
\label{Q35}
\end{equation}
As is obvious from Eq.(\ref{Q35}), the energy density of radiation, increases with time and tends to infinity at Big Rip singularity. This is because that with moving two branes towards each other, tachyon inflaton radiation grows and acquires the large values near the colliding point. Another interesting result that can be deduce from this equation is that the radiating energy density depends on the equation of state parameters and any enhancement or decrease in this density can be a signature of some interactions between two universes in extra dimension.

\section{Considering the signature of interaction between branes in observational data}\label{o2}
In previous section, we propose a model that allows to consider the warm inflation of universe in brane-antibrane system. Here, we examine the correctness of our model with observation data and obtain some important parameters like ripping time. Until now, eight possible asymptotic solutions for cosmological dynamics have been proposed
\cite{m25}. Three of these solutions have non-inflationary scale factor and another three ones
of solutions give de Sitter, intermediate and power-low inflationary expansions. Finally, two
cases of these solutions have asymptotic expansion with scale factor ($a = a_{0}exp(A[lnt]^{\lambda})$.
This version of inflation is named logamediate inflation \cite{m26}. Our model lets us to study
Warm-tachyon inflationary model in the scenarios of  logamediate
inflation with ${\lambda}=1$. In this condition, the model is converted to power-law inflation. Using (\ref{Q30}), the number of e-folds may be found:
\begin{equation}
N=\int_{t_{0}}^{t} H dt= |A|[ln\frac{t_{rip}-t_{0}}{t_{rip}-t}]
\label{Q36}
\end{equation}
where $t_{0}$ denotes the beginning time of inflation epoch.

In Fig.1, we present the number of e-folds \emph{N} for warm
scenario as a function of the t where t is the age of universe. In this plot, we choose A=-100, $t_{0}=0$ and $t_{rip}=33(Gyr)$. We find that \emph{N}=50 leads to $t_{universe}= 13.5(Gyr)$. This result is compatible with with both Planck and WMAP9 data \cite{m11,m12,m27}. It is clear that the number of e-folds \emph{N} is much larger for older universe. This is because, as the age of universe increases, the distance between branes becomes smaller and the tachyon field acquires bigger values.

\begin{figure}\epsfysize=10cm
{ \epsfbox{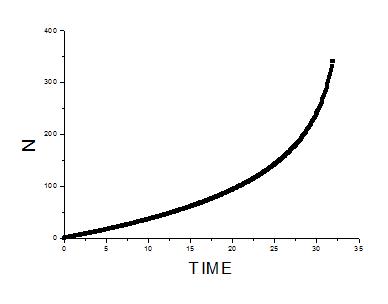}}\caption{The number of e-folds \emph{N} in warm inflation
scenario as a function of $t$ for A=-100, $t_{0}=0$ and $t_{rip}=33(Gyr)$. } \label{1}
\end{figure}

Another parameters that help us to test our model with experimental data are the power-spectrum of scalar and tensor perturbations which presented by \cite{m10,m28}:
\begin{eqnarray}
&&\Delta_{R}^{2}=- (\frac{\Gamma^{3}T^{2}}{36(4\pi)^{3}})^{\frac{1}{2}}\frac{H^{\frac{3}{2}}}{\dot{H}} \nonumber\\
&&\Delta_{T}^{2}= \frac{2H^{2}}{\pi^{2}}, \label{Q37}
\end{eqnarray}
where $T=[-\frac{r\dot{H}}{2C(1+\frac{r}{3}})]^{\frac{1}{4}}$, $r=\frac{\Gamma}{3H}$ and $C=\frac{\pi^{2}g^{\ast}}{30}$ ($g^{\ast}$ is the number of relativistic degree of freedom.). Using these parameters, we can define the tensor-scalar ratio as \cite{m10,m28}:
\begin{eqnarray}
&& R = -\frac{\Delta_{T}^{2}}{\Delta_{R}^{2}} = -(\frac{144(4\pi)^{3}}{\Gamma^{3}\pi^{4} T^{2}})^{\frac{1}{2}}\dot{H}H^{\frac{1}{2}}, \label{Q38}
\end{eqnarray}
In Fig.2, we show the tensor-scalar ratio \emph{R} for warm
scenario as a function of the age of universe. In this plot, we choose A=-100, $t_{0}=0$, c=.01 and $t_{rip}=33(Gyr)$. We find that \emph{R}=0.0147 leads to $t_{universe}= 13.5(Gyr)$. This result is compatible with with both Planck and WMAP9 data \cite{m11,m12,m27}. Obviously, the tensor-scalar ratio \emph{R} is much larger for older universe. The reason for this is that when two old brane universes  approach  together, the value
of tachyon inflaton increases and tends to infinity at singularity.

\begin{figure}\epsfysize=10cm
{ \epsfbox{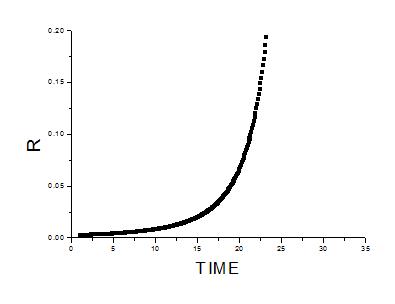}}\caption{The tensor-scalar ratio \emph{R} in warm inflation
scenario as a function of $t$ for A=-100, $t_{0}=0$, C=.01 and $t_{rip}=33(Gyr)$. } \label{2}
\end{figure}

Finally, we compare our model with the Scalar spectral index which is defined by\cite{m10,m28}:
\begin{equation}
n_{s}-1=-\frac{d ln\Delta_{R}^{2}}{d ln k}=\frac{3}{2}\varepsilon - (\frac{r'}{4r})\eta
\label{Q39}
\end{equation}
Here  k is co-moving wavenumber and $\varepsilon$ and $\eta$ are slow-roll parameters of the warm  inflation which are given by:
\begin{eqnarray}
&& \varepsilon =- \frac{1}{H}\frac{d ln H}{dt} \nonumber\\
&& \eta = -\frac{\ddot{H}}{H\dot{H}}, \label{Q37}
\end{eqnarray}
In Fig.3, we show the Scalar spectral index $\emph{n}_{s}$ for warm
scenario as a function of the  age of universe. In this plot, we choose A=-100 and $t_{0}=0$. Comparing this figure with figures(2,3), we find that $N\simeq 50$ case leads to $n_{s}\simeq 0.96$. This standard case is found in $0.01 < R_{Tensor-scalar } < 0.22$, which is consistent with both Planck and WMAP9 data \cite{m11,m12}. At this point, the finite time that Big Rip singularity occurs is $t_{rip}=33(Gyr)$.
\begin{figure}\epsfysize=10cm
{ \epsfbox{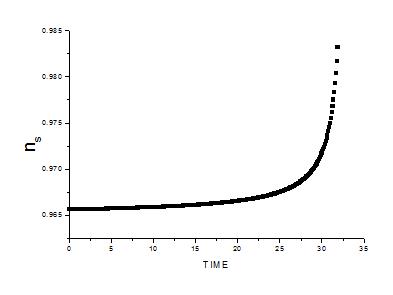}}\caption{The Scalar spectral index $\emph{n}_{s}$ in warm inflation
scenario as a function of $t$ for A=-100, $t_{0}=0$ and $t_{rip}=33(Gyr)$. } \label{3}
\end{figure}

\section{Summary and Discussion} \label{sum}
In this research,  we construct warm  inflation in brane-antibrane system and show that the energy density, slow-roll, Number of e-fold and
perturbation parameters can be given  in terms of the time that the two branes approached together in extra dimensions. According to our results, when the distance between branes  increases, the Number of e-fold, the spectral index and the tensor-scalar ratio  increases and tends to infinity at Big Rip singularity . This is because, as the  separation distance between branes decreases, the interaction potential increases and at higher energies there exists more channels for inflaton production  in brane-antibrane system; consequently, the effect of inflaton radiation from extra dimension on the universe inflation becomes systematically
more effective. We find that $N\simeq 50$ case leads to $n_{s}\simeq 0.96$. This standard case may be found in $0.01 < R_{Tensor-scalar } < 0.22$, which agrees with observational data \cite{m11,m12,m27}. At this point, the finite time that Big Rip singularity occurs is $t_{rip}=33(Gyr)$.

 \end{document}